\documentclass[runningheads]{llncs}
\usepackage{graphicx}
\usepackage[normalem]{ulem}

\begin{document}
\title{Ten AI Stepping Stones for Cybersecurity}
%
%
\author{Ricardo Morla
\authorrunning{R. Morla}
%
\institute{Faculty of Engineering and INESC TEC, University of Porto, Portugal} \\
\email{ricardo.morla@fe.up.pt}}
\maketitle              
\begin{abstract}
With the turmoil in cybersecurity and the mind-blowing advances in AI, it is only natural that cybersecurity practitioners consider further employing learning techniques to help secure their organizations and improve the efficiency of their security operation centers. But with great fears come great opportunities for both the good and the evil, and a myriad of bad deals. This paper discusses ten issues in cybersecurity that hopefully will make it easier for practitioners to ask detailed questions about what they want from an AI system in their cybersecurity operations. We draw on the state of the art to provide factual arguments for a discussion on well-established AI in cybersecurity issues, including the current scope of AI and its application to cybersecurity, the impact of privacy concerns on the cybersecurity data that can be collected and shared externally to the organization, how an AI decision can be explained to the person running the operations center, and the implications of the adversarial nature of cybersecurity in the learning techniques. We then discuss the use of AI by attackers on a level playing field including several issues in an AI battlefield, and an AI perspective on the old cat-and-mouse game including how the adversary may assess your AI power.

\keywords{Artificial Intelligence  \and Cybersecurity \and Battlefield.}
\end{abstract}

\section{What do you mean AI?}

Not the Peter F. Hamilton's Commonwealth Saga style of AI nor the equally brilliant dozen others like Asimov's I, Robot that thankfully populate our imagination. Not yet anyway. 
At the forefront of AI today are systems that learn how to perform a given task without having been specifically programmed for that task. These systems are able to learn the mappings from their inputs to their outputs, can learn more abstract representations of their inputs, and do this through multiple layers of input-to-output mapping, each layer with increasingly abstract representations. The term ``deep learning'' is used to acknowledge these multiple, stacked layers. In image processing, layers may provide a representation for increasingly abstract concepts of edges, contours, object parts, and objects that we want to identify. Have the AI learn from an adequate (large, diverse) data set of labeled cat and dog images, and it will likely perform well. Need to distinguish another pair of animals? No programming required, just feed another instance of the same AI with the data set of the new pair of animals and it will likely also perform well. As you can expect, a larger chunk of the work in using deep learning lies in preparing and feeding the data set rather than in actually developing and testing new models. Alternatively, you can have the AI learn from a reward function, for example wining or losing a game, using a technique called reinforcement learning~\cite{silver_mastering_2016}.
Many technical issues such as the number of layers and the type of network can have an impact on the performance of the AI too, which we will not discuss here. For a great conceptual overvirew of deep learning refer to chapter 1 in Goodfellow, Bengio, and Courville's Deep Learning book~\cite{Goodfellow-et-al-2016}, and continue reading the book for more technical details on deep learning.

\subsection{Other approaches to AI}

The deep learning approach to AI lies in the wider field of statistical machine learning, where you can find statistical models that can be learned from data. These include logistic regressions, support vector machines, component analysis, hidden Markov models, and (generically) probabilistic graphical models. For an in-depth technical description of many machine learning techniques refer to~\cite{hastie_elements_2009}. The major differences between these machine learning techniques and deep learning are 1) typically the set of features used in ML are chosen by the researcher rather than the algorithm and 2) the typically single layer approach in ML makes it much more complex to represent abstract concepts in the data. A very different approach to AI and in fact its classical and predominant approach in the mid-late 20th century uses rules to manipulate symbols and perform symbolic inference in an if-then logical approach. Expert systems with hand-crafted rules and well defined symbols were built using e.g. LISP and Prolog and applied in many specific tasks. For in-depth technical details on logic and knowledge-based systems refer to chapter 7 in~\cite{russell_artificial_2009}.

\subsection{Methodological concepts}

When going for machine learning type of AI we should be aware of some important methodological concepts. The training phase is typically an optimization process that searches for the best configuration of the AI for the given data, and it is where the AI learns how to perform its task. The training can be supervised if we provide labels to the training data -- for example the type of animal in the image -- or unsupervised if we want the AI to group pictures of the same type of animal together but not necessarily care about which type is which. Once the learning is over, we can use the AI to identify the type of animal in the inference phase. In some cases we may also want the AI to continue learning from new data during the inference phase. This process is called online, stream, or incremental learning. Although this may translate into a non-trivial technical issue of how to update the configuration of the AI, a more fundamental question is whether the underlying concepts that the AI is trying to learn changed significantly or not~\cite{gama_knowledge_2010}. Incremental learning allows the AI to track any drifts that may happen in the underlying concept and still consider the naturally occurring variance in the data. This is especially important in cybersecurity for tracking evolving normal usage, removing attack data from the learning of normal usage, and considering the possible adversarial nature of the learning data. Another important methodological concept is the independence of data samples from each other -- as for example is the case with the pictures of animals -- or if the data samples are somehow sequential as in the case of the images of a video. In that case the AI process may need to support memory, such that one or more of the previous data samples may be used both for learning and for inference. This can be the case for natural language processing and video~\cite{yue-hei_ng_beyond_2015}.
For insight on additional methodological concepts in machine learning such as the bias-variance tradeoff, model selection criteria,  the curse of dimensionality, and performance metrics for classification including accuracy, precision, and recall refer to chapter 1 in \cite{bishop_pattern_2006} and section II in  \cite{buczak_survey_2016}.

\subsection{Issues at the forefront of AI}

Moving towards AI singularity or not, the advances in AI lead to a number of issues that are both technical and societal~\cite{taylor2016alignment}: will the AI perform as intended by its operators, how can the AI fail? One way of moving forward here is to have the AI explain why it made a particular decision~\cite{DBLP:journals/corr/abs-1710-00794}. This is especially hard for deep learning with its multiple layers and complex interconnections. One may also question if the AI can be tricked into making wrong decisions~\cite{Papernot:2017:PBA:3052973.3053009}, either by slightly modifying the input data for AI decision or by poisoning the training data and altering the learning process. In fact, explainability and resilience to adversarial attacks are two of of the challenges identified by UC Berkeley~\cite{stoica_berkeley_2017} for mission critical AI usage (which includes cybersecurity), together with continuous learning and secure computing. Researchers have been questioning the use of learning from data in cybersecurity at least since Sommer and Paxson's seminal paper~\cite{sommer_outside_2010} on the weaknesses of using machine learning outside the closed world of intrusion detection research. This paper is a turning point for machine learning in cybersecurity as it reflects on issues such as closing the semantic gap of black box models and considering the adversarial nature of the attacker, and which in fact crosscut application domains in AI.

\section{How do I use it, then?}

That depends on what specifically you want to do in cybersecurity.

You may want to use an AI to help you detect or predict a malicious event. If you know the malicious event and can handwrite some rules that parse your data and signal the event, you don't really need an AI. This is the typical signature-based approach to virus detection. If you know the event, have access to a sufficiently large number of data samples where the event occurs, but the event seems too complex for hand-written rules, then you can use supervised learning. Take the data samples from the malicious event and from the normal events, label them, and feed the data and labels to the AI. You will need to ponder on the data representation and features, but more of that later. The major problem is that if a new, previously unseen malicious event is different from your known malicious events then you will likely not be able to detect it. If you don't know the event then you have to resort to anomaly detection, which is basically saying that you want the AI learn what the ``normal'', non-malicious behavior is. Set a threshold for the malicious event and you're ready to go. Because the AI doesn't have samples for the malicious events, the major difficulty with this approach is figuring out the level of detail with which the ``normal'' behavior should be modeled.  
AI empowered detection can be of use in many of the kill chain phases, by helping to identify e.g. the delivery of a weaponized document, the exploitation of vulnerabilities, the traffic used for command and control, and the exfiltration of sensitive data.
Malicious event detection is likely the best-known use of AI in cybersecurity, but definitely not the only. You may also want to: 2) find causes of malicious events by correlating with data samples from other sources, 3) find similar malicious events by defining a distance metric between them and grouping according to that distance, 4) identify patterns of attacks by finding frequently occurring sequences of malicious events. You can go beyond malicious events and consider using AI to rank, filter, organize, and interpret cybersecurity intelligence, and to prioritize and contextualize vulnerability scans and incident report response, just to mention a few. 

\subsection{Learning and inference}

When and where does the AI learn about malicious events, normal behavior, and any other cybersecurity-relevant piece of information? Not necessarily at the same time and location where it will take a decision. In a self-driving fleet of cars, for example, driving decisions need to be taken within a sub-second delay in every car, but learning may take place at a central location and requires sufficiently long time for the images and sensor data relayed by the cars to be filtered, labeled, and fed into the AI. 
The newly trained AI can then be downloaded to the cars, somehow similarly to updating your anti-virus definitions daily. This approach would allow the cybersecurity AI to evolve and possibly support new cybersecurity threats learned at a central location, even possibly adjusting for local behavior if local data can be sent to the central learning location. Alternatively, you may allow at least some part of your AI to continuously learn from your local cybersecurity data sources for more quickly adjusting to what's happening locally. How fast does a cybersecurity AI need to learn new concepts? Assuming that you want to have an AI decision as fast as possible, then the AI should be able to learn a new attack as fast as possible. This could be the detection of a new malicious pattern or of a new ``normal'' behavior in your organization. A related question is how do you adjust the mix between local knowledge and the more global knowledge that may be available from a central location? Assuming you download the newly trained global AI from the central location, one option would be to have the two AIs running in parallel and decide on an approach at arbitration when they don't agree. Another option would be to train the local instance of the global AI with a replay of the local data up to that point, leading to an AI that has both the global and local knowledge and that could continue to learn from local data as it arrives.

\subsection{Data sources}

Buczak et al. \cite{buczak_survey_2016} provide a description of the packet and flow data sources typically used in network intrusion detection. Flow data is a summary of packet data for each flow, containing e.g. the number of packets and the total number of bytes in each direction of the flow, typically with a much larger time constant than packet data. Depending on the task, feeding packet size, time, and direction directly to the AI may be more useful than summarized flow data~\cite{Bernaille:2006:TCF:1129582.1129589}. Although this seems to suggest that the more data the better, providing raw packets to the AI may not always be the best approach. Network traffic is highly structured with much information standardized and packaged according to Internet standards like TCP/IP. This brings about a point of what the deep learning approach to AI is best at: learning ``intuitive'' abstractions from the data~\cite{Goodfellow-et-al-2016} which are difficult or too complex for a human to code. Using existing, human coded TCP/IP protocol parsers such as Wireshark and feeding the output of those parsers to the AI may be a better option in general ~\cite{10.1007/978-3-319-70139-4_88}. Having said this, in some very specific cases where the Internet stack is not broken but  otherwise abused by attackers in order e.g. to exfiltrate information, it may be interesting to directly feed raw packet data to the AI, much as we feed raw pixel data to an image processing AI. This kind of discussion may also apply to other data sources  typically used in cybersecurity, including the output of rule-based network intrusion detection systems like Suricata, of host based intrusion detection systems like OSSEC, and of local or centralized anti-virus solutions, the logs of applications, servers, and network services and devices, and incident reports. In addition to this data that originates from within the organization, a large collection of open source intelligence data is available from which an AI could learn.

\subsection{Infrastructure}

Finding the right data source can be a difficulty. Plumbing that data from different places in your organization all the way to the AI -- for training or inference or both -- will definitely be another. Restrictions on the available bandwidth or the amount of data that can be stored may have an impact on the choice of data to feed the AI. When upgrading to AI in the security operations center SOC, the most straightforward approach will be to use whatever data is available in the security information and event management SIEM or in other components in the SOC. While this may be interesting for quick wins, small-scale experiments may help show the value of a given data source and support investment decisions to make new data available in the SOC. A variety of processing, storage, and networking hardware can be used for the AI together with software stacks for big data and deep learning. How to achieve desired performance with minimum hardware cost is an important issue here, as is comparing the performance of different deep learning software options. Refer to~\cite{hazelwood_applied_2018} for an example of a machine learning infrastructure at Facebook, and to~\cite{shi_benchmarking_2016} for an example of benchmarking different software stacks for deep learning.

\section{Ok, but I don't really need it.}

Despite the current AI hype, learning from data is not new to cybersecurity and has been a driver of research on intrusion detection systems at least since the early nineties. Part of that research has found its way into commercial and open source products and has been accounted for in numerous surveys~\cite{buczak_survey_2016}, \cite{tsai_intrusion_2009}. AI has extended to areas beyond intrusion detection. If you're running a security operations center and relying on some sort of anti-virus, spam detection engine, or advanced SIEM, you may already be using it. One particularly interesting example is malicious software: malicious software binaries were traditionally identified by unique sequences of bytes seen as ``signatures'' of the malware. Today we have an exploding number of signatures and malware generation tools can dynamically create different byte codes and signatures for the same malware. This opens the door to reverse engineering approaches that look at system calls, memory and file system access, interaction with operating system registry and other properties of the executable to identify malware. Given the diversity of features and the variability of the values they can take, handwriting detection rules is hard. Consequently, AI is starting to be used to learn and detect new malware from data samples, identify malware variants, learn which category the malware falls under, and find similarities and novelty in new malware. For a detailed discussion and review of related work on AI for malware refer to~\cite{ucci_survey_2017}. In table \ref{long} we highlight several domains in cybersecurity -- mostly under operations or intelligence -- where AI has been applied, and identify an example contribution in each domain. Other domains\footnote{\url{https://www.linkedin.com/pulse/map-cybersecurity-domains-version-20-henry-jiang-ciso-cissp/}} may benefit from AI, for example the use of AI in penetration testing from an attacker's perspective, and the use of AI in game-theory and reinforcement-learning-like active network defense situations. A hierarchical approach to the AI may  be viable in the SOC, where output from specific AI tasks such as host intrusion detection and exfiltration detection are provided as input to an upper layer AI. This AI would be able to learn from the variety of lower-layer AI outputs, having an image-like view of the organization and the ability to abstract higher level goals and keeping the organization safe from attacks. One might ponder what similarity this would have with actual images and image processing AIs at the level of the deep learning network architecture. An alternative approach to this layered architecture is to encompass all the lower-layer cybersecurity tasks into a single deep learning solution, in the expectation that the abstraction power of the deep layers would outperform the hierarchical approach and assuming enough computation power and data sets are available to train the single solution.

\begin{table}
\begin{center}
\begin{tabular}{ l | l}
\hline
Domain & Approach \\
\hline
\hline
Malware~\cite{ucci_survey_2017} & Detect, find similarities; use system calls, etc.\\ 
\hline
Phishing web sites~\cite{sahoo_malicious_2017} &  Detect; use static and sandboxed dynamic content \\ 
\hline
Spam~\cite{blanzieri_survey_2008} & Detect; use email content and headers\\
\hline
Host intrusion~\cite{kim_lstm-based_2016} & Detect; use system call natural language modeling\\
\hline
Network intrusion~\cite{zhang2001hide} & Detect in hierarchy; use traffic and anomalies from sub-layers\\
\hline
Hardware intrusion~\cite{ren_learning-based_2016} & Detect; from JTAG instructions\\
\hline
User behavior~\cite{KENT2015150} & Model; use authentication graph\\
\hline
Exfiltration~\cite{das_detection_2017} & Detect; use DNS URL\\
\hline 
0-day~\cite{sun_using_2018} & Detect; use attack graph of non-zero day intrusions\\
\hline
DDoS~\cite{doshi_machine_2018} & Detect; use specific traffic features for IoT botnet\\
\hline
Intelligence feed~\cite{meier_feedrank:_2018} & Rank feeds; use feed IoC-defined dependency and graph\\
\hline
IoC~\cite{gascon_mining_2017} & Find Indicator of Compromise similarities; use  IoC content \\
\hline
Incident response~\cite{woods_data_2015} & Find similar incidents; use IoC for similarity\\
\hline
TTPs~\cite{husari_ttpdrill:_2017} & Extract; use unstructured text and map to kill chain\\
\hline
\end{tabular}
\caption{Use of AI in example cybersecurity tasks, necessarily ongoing. \label{long}}
\end{center}
\end{table}




\section{What about all those privacy \sout{nutters} concerned people and laws?}

There's no way around it: privacy is here for the long run and cybersecurity AI must oblige. This has consequences in the amount and type of data that a security operations center can use for their AI. Add to the natural unease of sharing potentially sensitive business information with others -- be it companies or governments, and you're a step away from freezing the defender and making the attacker's day. What you need are tools that allow you to control and monitor the amount of information you remove from your data before you can use it in your AI or share it with others, while at the same time making sure that the AI can still learn something useful for the cybersecurity task at hand. Significant progress has been made in privacy-preserving data mining. Major approaches include applying some level of randomization to the data before data is delivered to the AI; changing values to a more general representation (e.g. integers to intervals); or removing some of the data values (sample- or field-wise). Notable privacy-preserting techniques include k-anonymity where any data sample is made indistinguishable from at least k-1 other samples, and the $\epsilon$-differential privacy where the impact of removing a single record from a data set -- and thus potentially revealing private information -- is negligible up to $\epsilon$.  Approaches to enable sharing with privacy can rely on cryptographically-based secure computing techniques or on distributed implementations of learning algorithms that share only learned parameters and not actual data. For an in-depth review of privacy-preserving techniques and metrics refer to~\cite{mendes_privacy-preserving_2017}. 

During the process of privacy regulation conformity you may end up defining which private data you use and for how long. Having techniques for properly anonymizing data and measuring their impact will help you not only better protect the privacy of your constituency but to deliver a better argument to your privacy protection officer. Cryptography-based solutions for sharing e.g. IoCs have been proposed~\cite{van_de_kamp_private_2016}, however these are for relatively simple tasks of counting sightings of IoCs. For a review on sharing in intrusion detection refer to~\cite{vasilomanolakis_taxonomy_2015}. Sharing for more advanced cybersecurity AIs will likely require more complex cryptography-based mechanisms or distributed learning where the learned parameters do not reveal private or business-sensitive information.

\section{Anyway, I don't trust it.}

And you're probably right not to, at least not without some help. In fact, the advantage of deep learning to intuitively capture the complexity of the problem is also one of its shortcomings. So how do you trust this system? You need to interact with it, test it with different samples, and understand its decisions. You can then improve it by providing additional data that you expect will make the learning more comprehensive. A quick search for deep learning-based intrusion detection, malware detection, and threat detection reveals that most studies only focus on the accuracy of the results without giving much thought to understanding the decisions or the semantics of the model. Comparing these studies with what is typically done in the image processing area, we come to two conclusions: 1) small to none effort to explain the rationale behind the specific deep learning architecture used, the choice of input for the first layer, and the number of layers; 2) no effort to understand the semantics of the abstraction that each layer provides. Take the image processing example in \cite{zeiler_visualizing_2013}: the first layer shows edges, the second contours, the third object parts, and finally the last layer takes object parts and combines them into objects. The neural network hidden ``units'' in each layer encode different concepts and can be ``activated'' onto the lower layers, creating synthetic representations of what that unit has learned and that we can visualize. 
Visually representing images is easier than network and system data of course, but that doesn't mean we cannot try. What could be the layers of a system that tries to learn normal behavior in an organization network and system? Communication patterns for HTTP, SSH, SMTP at one level? Email, web browsing, voice call, backup application behavior at the next level? Admin, engineering, accounting user behavior at the top? What if the system learns from attack samples rather than normal behavior? Does the AI learn this by itself, and can we check?
Cybersecurity can have complex data samples such as a comprehensive snapshot of the organization network or the malware features obtained by reverse-engineering binaries. 
Taking the image analogy from~\cite{zeiler_visualizing_2013} a step further into understanding a cybersecurity AI, we can ask what it would mean to check for ``feature invariance'' like image translation, rotation, and scaling in malware detection data, and to check the impact of zeroing-out part of the data sample in intrusion detection. Figuring out the equivalent of typical image-processing concepts in cybersecurity is an obvious path towards better explaining and trusting AI in cybersecurity. One might also contend that because of the diversity of tasks and data, new ways to validate and explain an AI can be developed based on specific requirements of cybersecurity.

\section{And it can be tricked.}

Again, you're probably right. Tricking an AI has been done many times, and adversarial learning is a focus of research in image processing and elsewhere. The main issue with the adversarial mindset is that machine learning techniques expect some form of random distribution to their input data, whereas an adversary will often be smarter than a random distribution. Think about it as the standoff between error detection mechanisms and cryptographic integrity in communication networks; errors are expected to be random, while attacks to integrity can target specific vulnerabilities of the cryptographic mechanisms. Adversaries can attack the AI's integrity by either crafting a data sample that looks normal to humans but that is misclassified by the AI during the inference phase (called an exploratory attack) or by adding, removing, or changing data samples in the training phase (called a causative or poisoning attack) so that the AI is not able to adequately learn the intented concepts. Causative attacks assume flaws in the AI training production chain that provide the attacker with write access to the training data set or, maybe more plausible, that the AI continues learning during the inference phase with live data from the organization's network on which the attacker has a compromised machine that can generate traffic. The complexity of a deep learning AI makes it non-trivial to understand how to come up with adversarial data samples that can be used for exploratory or causative attacks, and as it is typical in most attacks, the process of creating attack data samples can benefit from inside knowledge on the training data, the deep learning network architecture, and even from samples or oracle-like access to input/classification results. For a discussion on threat models, a review of adversarial attacks to non-deep learning machine learning, and an approach for crafting adversarial inputs for deep learning refer to~\cite{papernot_limitations_2016}. Defensive techniques do exist that can offset the attacker to an AI. Data sets should be cleansed, the impact of new samples analyzed before feeding them to the AI learning, and algorithms should be made more robust. For a detailed discussion on several machine learning defense techniques refer to~\cite{liu_survey_2018}. In any case, understanding the AI -- which intermediate concepts it is able to abstract, how resilient it is to feature invariance and to hiding fields in the data seem -- will likely only help to make it more resilient to adversarial attacks. 
If you're looking for examples of adversarial attacks to deep learning intrusion detection systems and malware detection, start here: \cite{madani_robustness_2018}~assumes a continuous learning autoencoder-based intrusion detection system and assesses its robustness to poisoning attacks under potential, naturally-occurring, non-malicious concept drifts; \cite{al-dujaili_adversarial_2018}~evaluates the effectiveness of several adversarial methods on the detection of a set of malware Portable Executable files.

\section{Are you telling me the bad guys can use it as well?}

Definitely yes. In addition to the adversarial learning techniques discussed in~\cite{papernot_limitations_2016}, an attacker can resort to generative models and try to learn what 'normal' data looks like for a given detector and generate malicious samples that look like normal and go unnoticed through that detector. Generative techniques range from statistical random number generators such as mixtures of Gaussians, hidden Markov models, and other probabilistic graphical models, to more recent deep learning techinques including autoencoders and generative adversarial networks. For more details on generative models and their evaluation see~\cite{theis_note_2015}. On the flip side, you can have attackers using AI to perform side-channel attacks to privacy by inferring user behavior from eavesdropped traffic with the help of discriminative machine learning techniques. Using AI is also becoming more practical; deep learning platforms are increasingly providing support for browser and mobile apps~\cite{xu_first_2018} which opens the door for generative-based malware C2 traffic communications~\cite{rigaki_bringing_2018} and generative-based DGA (dynamically-generated domain names~\cite{anderson_deepdga:_2016} among others. 

If we can disentangle ourselves for just a moment from the good guy vs. bad guy view of the world, the duality of the problem becomes clearer. The defender wants to know what the attacker is doing (e.g. detecting intrusions) and also to prevent the attacker from knowing what it is doing (e.g. for privacy). The attacker wants the same thing: to eavesdrop on the defender and to avoid being detected. So at an abstract level we can envision a pair of opponent AIs where the samples of the generative AI are designed to break the discriminatory AI and the discriminative AI learns to distinguish between normal samples and samples from the generative AI -- regardless of which side (good vs. bad) we are currently looking from. In the next sections we discuss some of the rationale in this battle of AIs approach to cybersecurity, provide examples of AI cybersecurity battlefields, and discuss how opponents can gather information from their counterpart AI through Intelligence and reverse engineering techniques.

\section{So what does a battle of AIs look like?}

\subsection{Level playing field}

Bringing in additional and more detailed features to one side of the battle only is likely to disrupt the balance between attacker and defender. This is the case for example in malware C2 traffic when the attacker has learned to generate tuples of flow size and duration that resemble those of normal traffic, but uses a hardwired process for generating the actual packets of the flow. While the generated flow has duration and size that match the normal traffic, more detailed features such as the packet size and inter-packet time distribution may not. Bringing these more detailed features in allows the defender to outsmart the attacker no matter how good the attacker's generator of flow size and duration is. Unless there is very strong evidence of which level of features the opponent uses, trying more detailed features seems to make sense. One question that we can ask is if we can figure out which features the opponent uses. This can be related to black-box attacks that try to infer which data is used for training or in the case of image processing how much scaling is done on the image before actual AI processing. 


The other relevant question is if there is a set of features detailed enough for an attacker to be sure the detector cannot use more detailed features. And the answer is yes -- those features are the actual data. This of course depends on the battlefield: for example for domain names this would be the actual domain name rather than n-gram statistics, and for malware C2 traffic this would be the actual packet contents and timings, although with encryption the only relevant packet contents are those not encrypted in the different headers. Using the actual data rather than features is quickly becoming the norm with deep learning image processing leading the way. If this can be done in a given battlefield is a very relevant question, which needs to consider not only modeling capability but also sustaining possibly high bit rates. If so, using the full data levels the playing field between attacker and defender AIs. It also brings us to a set of relevant questions related to the extent to which data morphing is possible for a given battlefield and to how good an AI must be to avoid detection or detect traffic from another AI.

\subsection{Mine beats yours}

How far can we go into understanding how good an AI must be to beat its opponent? Let's try to get some intuition from an example. Assume a battlefield of independent samples of real-valued data, in which opponents have to generate and make a decision for each sample independently -- so no order or correlation is used in the generation and detection of samples. Now think of a simple outlier detector that blocks samples that are outside of the range $\mu \pm 3 \sigma$. The attacker wants to generate samples that are not blocked by the detector. What does the attacker know about the detector? 1. If the attacker knows the model and its estimated parameters $\mu$ and $\sigma$, then it can randomly generate samples that are within the $\mu \pm 3 \sigma$ range and successfully perform the attack on every sample. 2. If the attacker knows the model, does not know the parameters, but can have access to training data, it can try to estimate the parameters from the data itself. The success of this attack will depend on how similar the attacker data is from the data used to train the detector, and on the randomness of the training algorithm; for example in the case of K-Means, mixture of Gaussians models, or neural networks in general, the training depends on random initialization values. 3. If the attacker does not know the model, then it is left with guessing which model the detector uses. As in any guess, some uncertainty is involved and it makes sense to try to understand the impact of this uncertainty. Let's say the detector has been upgraded to use a mixture of Gaussians and the generator still uses the $\mu \pm 3 \sigma$ model, inferred from the same training data as the detector. With luck, some of the generated samples will fall within one of the Gaussians of the detector, but not all; how many will depend on the data and on the number of Gaussians used in the mixture.  If the generator model was upgraded to a mixture of Gaussians but not the detector, intuitively the generated samples would better fit the data and possibly fall within the $\mu \pm 3 \sigma$ range -- but not necessarily. So a mismatch of models may benefit one of the opponents. Fast forward to deep learning and you can break down the uncertainty of the models of the opponent by the number of layers, type of neural network, reuse of part of the models known as transfer learning, and many other systematizations of deep learning.

\subsection{Full morphing ability}

When generating adversarial samples, one question that is very relevant is if for the given battlefield the relation between generated sample shape and intended behavior is somehow constrained. Can the generator take an intended adversarial behavior sample, morph it into any other that is not detected, and keep its adversarial behavior -- or is it somehow constrained? For example, an attacker that wants to send terror-related emails without being detected has to write a text that is not only intelligible but maintains its subversive message while at the same time not being picked up by the email detector AI. This constraint on the morphing ability could have an impact on how the opponents build their discriminative and generative systems. Differently, it could happen that a given battlefield has full morphing ability, meaning that the generator can choose whatever shape it wants for the adversarial sample. The best example here is likely eavesdropping on encrypted traffic. As long as the generator is able to mimic the non-encrypted part of the communication including timings, sizes, and non-encrypted headers like TLS fingerprints~\cite{anderson_deciphering_2018}, and assuming the detector is not able to break the encryption, the encrypted channel will provide cover to whatever adversarial plaintext data sent on the encrypted payload. In this case, one concern for the generator is if the morphed encrypted traffic is able to sustain the requirements of the adversarial behavior, for example related to throughput and latency.


\subsection{Winning with single adversarial sample}

If the generator can find one sample that is classified as normal by the detector then why not use it all the time? If the detector does not keep some memory of prior samples, this would work. Repeating the exact same sample would be straightforward to detect leading to sample blacklist. Slightly changing a Gaussian sample might work, but the more complex the sample domain the harder it will be for the attacker to estimate how much and what can be changed. So regardless of the complexity of the sample, you always want to check if the sample is repeated. You could also check for well-known normal samples from public databases. One venue for attack would be to take these known samples and check to which extent they can be changed to be detected as normal but not as repeated.


\section{Example battlefields}

\begin{itemize}

\item \textbf{Malware C2 Traffic}. Command and control traffic from compromised hosts has evolved to make detection harder -- e.g. by moving from IP addresses to dynamically generated domain names and by using encryption. At the same time detectors also evolve to include more detailed features. Morphing C2 traffic to avoid detection has been shown in~\cite{rigaki_bringing_2018} with the help of Generative Adversarial Networks. In this battle, level playing field equates to encryption and full morphing ability is possible with C2 traffic requirements constraints.

\item \textbf{Traffic Privacy}. Encryption has stepped up in web, smart home, mobile apps, voice, and video traffic across the Internet to protect users against eavesdroppers. With privacy tools like TOR, it is harder to find server IP addresses and infer behavior from encrypted communication patterns. However, most communication today is not over TOR and privacy attacks have been the subject of extended research~\cite{monaco_what_2019} including traffic morphing techniques~\cite{luo_httpos:_2011}. This battlefield is very similar to malware C2 traffic with good guy vs. bad guy roles switched. One difference is that malware will try harder to hide in the shadow of normal traffic while users concerned with privacy will rather be more worried about the attackers finding out what they're doing and which sites they are visiting so a constant bitrate approach may work for them as long as the quality of service is not compromised. 

\item \textbf{Malware domain names}. Domain generation algorithms (DGAs) are used to generate pseudo-random domain names that are easy to generate by both the malware and the command and control server and difficult to detect by security operators. Researchers have applied unsupervised machine learning to train a domain name generator from legitimate domain names~\cite{anderson_deepdga:_2016}. Although their aim was to improve existing detection algorithms and not to propose a new malware, their results show how close generated and legitimate domain names can be. With the availability of deep learning engines for mobile devices and browser applications, it is reasonable to expect that malware will sooner or later leverage machine learning algorithms. The full data is the actual domain name, and generators have full morphing ability except for collisions with already registered domain names; however, collisions are typically taken care of at a later stage.

\item \textbf{Executable Detection}. Finding an executable with a blacklisted hash was for a long time the way to detect hosts to which malware had been downloaded. As always, attackers adapted and started to slightly change their binaries yielding very different hash values while keeping the functionality. Static and dynamic analysis of malware opened the door to machine learning approaches to detect malware executables, which again prompted the development of adversarial learning techniques on the malware~\cite{kolosnjaji_adversarial_2018}. The full data for this battlefield is the malware binary and generators are limited to a viable binary that does not change the intended malicious behavior.

\item \textbf{Browser Document Object Model}. Malicious browser extensions can be used to perform a variety of attacks inside the browser. Malicious browser extensions can be detected through static analysis, which is much similar to the malware executables battlefield~\cite{dekoven_malicious_2017}.  Additionally, as malicious extensions change application behavior through erroneous DOM mutations, application monitoring code can check if DOM mutations are consistent with application behavior. The complexity of the DOM structure and the intense activity in the browser is likely to call for learning from the data both for malicious DOM mutation detection and detection avoidance. The full data here are the sequences of DOM mutations and full morphing ability is constrained by the actual behavior of the DOM mutations.

\end{itemize}

\section{Can you see my AI?}

If the AI is going to be used in the context of cybersecurity operations  with enough resources to gather intelligence data and profile its potential attackers, and if the threat model for adversarial attacks depends on the capabilities of the AI like the specifics of the architecture of the deep learning network and access to training data samples, then it is foremost important to ask yourself how attackers can gain that kind of information. In addition to the old spy game, the open source data policies to which government organizations and public corporations are typically vulnerable to because of transparency concerns can disclose much information. Moreover, the daily use of the AI itself may be more revealing than desired. A variety of techniques for reverse engineering AIs have been proposed in the literature, from side-channel attacks leveraging shared resources to learn DNN architectures~\cite{yan_cache_2018} to queries to the API to extract a surrogate model~\cite{juuti_prada:_2018}, and  membership inference of training data~\cite{hayes_logan:_2017}.  Practical attacks to deep neural networks built on top of public models (named transfer learning)~\cite{wang_great_2018} point the way to further assumptions and inference attacks on an AI. \cite{juuti_prada:_2018} also discusses an approach to detect API queries attacks and points to how you could hide the details of your AI.


\bibliographystyle{splncs04}
\bibliography{tensteppingstones}

\begin{thebibliography}{10}
\providecommand{\url}[1]{\texttt{#1}}
\providecommand{\urlprefix}{URL }
\providecommand{\doi}[1]{https://doi.org/#1}

\bibitem{al-dujaili_adversarial_2018}
Al-Dujaili, A., Huang, A., Hemberg, E., O'Reilly, U.M.: Adversarial {Deep}
  {Learning} for {Robust} {Detection} of {Binary} {Encoded} {Malware}.
  arXiv:1801.02950 [cs, stat]  (Jan 2018),
  \url{http://arxiv.org/abs/1801.02950}, arXiv: 1801.02950

\bibitem{anderson_deciphering_2018}
Anderson, B., Paul, S., McGrew, D.: Deciphering malware’s use of {TLS}
  (without decryption). Journal of Computer Virology and Hacking Techniques
  \textbf{14}(3),  195--211 (Aug 2018). \doi{10.1007/s11416-017-0306-6},
  \url{http://link.springer.com/10.1007/s11416-017-0306-6}

\bibitem{anderson_deepdga:_2016}
Anderson, H.S., Woodbridge, J., Filar, B.: {DeepDGA}: {Adversarially}-{Tuned}
  {Domain} {Generation} and {Detection}. In: Proceedings of the 2016 {ACM}
  {Workshop} on {Artificial} {Intelligence} and {Security}. pp. 13--21. {AISec}
  '16, ACM, New York, NY, USA (2016). \doi{10.1145/2996758.2996767},
  \url{http://doi.acm.org/10.1145/2996758.2996767}

\bibitem{Bernaille:2006:TCF:1129582.1129589}
Bernaille, L., Teixeira, R., Akodkenou, I., Soule, A., Salamatian, K.: Traffic
  classification on the fly. SIGCOMM Comput. Commun. Rev.  \textbf{36}(2),
  23--26 (Apr 2006). \doi{10.1145/1129582.1129589},
  \url{http://doi.acm.org/10.1145/1129582.1129589}

\bibitem{bishop_pattern_2006}
Bishop, C.: Pattern {Recognition} and {Machine} {Learning}. Information
  {Science} and {Statistics}, Springer-Verlag, New York (2006),
  \url{//www.springer.com/gp/book/9780387310732}

\bibitem{blanzieri_survey_2008}
Blanzieri, E., Bryl, A.: A survey of learning-based techniques of email spam
  filtering. Artificial Intelligence Review  \textbf{29}(1),  63--92 (Mar
  2008). \doi{10.1007/s10462-009-9109-6},
  \url{https://link.springer.com/article/10.1007/s10462-009-9109-6}

\bibitem{buczak_survey_2016}
Buczak, A.L., Guven, E.: A {Survey} of {Data} {Mining} and {Machine} {Learning}
  {Methods} for {Cyber} {Security} {Intrusion} {Detection}. IEEE Communications
  Surveys Tutorials  \textbf{18}(2),  1153--1176 (2016).
  \doi{10.1109/COMST.2015.2494502}

\bibitem{das_detection_2017}
Das, A., Shen, M.Y., Shashanka, M., Wang, J.: Detection of {Exfiltration} and
  {Tunneling} over {DNS}. In: 2017 16th {IEEE} {International} {Conference} on
  {Machine} {Learning} and {Applications} ({ICMLA}). pp. 737--742 (Dec 2017).
  \doi{10.1109/ICMLA.2017.00-71}

\bibitem{dekoven_malicious_2017}
DeKoven, L.F., Savage, S., Voelker, G.M., Leontiadis, N.: Malicious browser
  extensions at scale: {Bridging} the observability gap between web site and
  browser. In: 10th \$\{\${USENIX}\$\}\$ {Workshop} on {Cyber} {Security}
  {Experimentation} and {Test} (\$\{\${CSET}\$\}\$ 17).
  USENIX\$\vphantom{\{}\}\$ Association\$\vphantom{\{}\}\$ (2017)

\bibitem{DBLP:journals/corr/abs-1710-00794}
Doran, D., Schulz, S., Besold, T.R.: What does explainable {AI} really mean?
  {A} new conceptualization of perspectives. CoRR  \textbf{abs/1710.00794}
  (2017), \url{http://arxiv.org/abs/1710.00794}

\bibitem{doshi_machine_2018}
Doshi, R., Apthorpe, N., Feamster, N.: Machine {Learning} {DDoS} {Detection}
  for {Consumer} {Internet} of {Things} {Devices}. arXiv:1804.04159 [cs]  (Apr
  2018), \url{http://arxiv.org/abs/1804.04159}, arXiv: 1804.04159

\bibitem{10.1007/978-3-319-70139-4_88}
Fang, P., Huang, L., Zhang, X., Xu, H., Wang, S.: Detect malicious attacks from
  entire tcp communication process. In: Liu, D., Xie, S., Li, Y., Zhao, D.,
  El-Alfy, E.S.M. (eds.) Neural Information Processing. pp. 867--877. Springer
  International Publishing, Cham (2017)

\bibitem{gama_knowledge_2010}
Gama, J.: Knowledge {Discovery} from {Data} {Streams} (May 2010),
  \url{https://www.crcpress.com/Knowledge-Discovery-from-Data-Streams/Gama/p/book/9781439826119}

\bibitem{gascon_mining_2017}
Gascon, H., Grobauer, B., Schreck, T., Rist, L., Arp, D., Rieck, K.: Mining
  {Attributed} {Graphs} for {Threat} {Intelligence}. In: Proceedings of the
  {Seventh} {ACM} on {Conference} on {Data} and {Application} {Security} and
  {Privacy}. pp. 15--22. {CODASPY} '17, ACM, New York, NY, USA (2017).
  \doi{10.1145/3029806.3029811},
  \url{http://doi.acm.org/10.1145/3029806.3029811}

\bibitem{Goodfellow-et-al-2016}
Goodfellow, I., Bengio, Y., Courville, A.: Deep Learning. MIT Press (2016)

\bibitem{hastie_elements_2009}
Hastie, T., Tibshirani, R., Friedman, J.: The {Elements} of {Statistical}
  {Learning}: {Data} {Mining}, {Inference}, and {Prediction}, {Second}
  {Edition}. Springer-Verlag, New York, 2 edn. (2009)

\bibitem{hayes_logan:_2017}
Hayes, J., Melis, L., Danezis, G., De~Cristofaro, E.: {LOGAN}: {Membership}
  {Inference} {Attacks} {Against} {Generative} {Models}. arXiv:1705.07663 [cs]
  (May 2017), \url{http://arxiv.org/abs/1705.07663}, arXiv: 1705.07663

\bibitem{hazelwood_applied_2018}
Hazelwood, K., Bird, S., Brooks, D., Chintala, S., Diril, U., Dzhulgakov, D.,
  Fawzy, M., Jia, B., Jia, Y., Kalro, A., Law, J., Lee, K., Lu, J., Noordhuis,
  P., Smelyanskiy, M., Xiong, L., Wang, X.: Applied {Machine} {Learning} at
  {Facebook}: {A} {Datacenter} {Infrastructure} {Perspective}. In: 2018 {IEEE}
  {International} {Symposium} on {High} {Performance} {Computer} {Architecture}
  ({HPCA}). pp. 620--629 (Feb 2018). \doi{10.1109/HPCA.2018.00059}

\bibitem{husari_ttpdrill:_2017}
Husari, G., Al-Shaer, E., Ahmed, M., Chu, B., Niu, X.: {TTPDrill}: {Automatic}
  and {Accurate} {Extraction} of {Threat} {Actions} from {Unstructured} {Text}
  of {CTI} {Sources}. In: Proceedings of the 33rd {Annual} {Computer}
  {Security} {Applications} {Conference}. {ACSAC} 2017, ACM, New York, NY, USA.
  \doi{10.1145/3134600.3134646},
  \url{http://doi.acm.org/10.1145/3134600.3134646}

\bibitem{juuti_prada:_2018}
Juuti, M., Szyller, S., Marchal, S., Asokan, N.: {PRADA}: {Protecting} against
  {DNN} {Model} {Stealing} {Attacks}. arXiv:1805.02628 [cs]  (May 2018),
  \url{http://arxiv.org/abs/1805.02628}, arXiv: 1805.02628

\bibitem{van_de_kamp_private_2016}
van~de Kamp, T., Peter, A., Everts, M.H., Jonker, W.: Private {Sharing} of
  {IOCs} and {Sightings}. In: Proceedings of the 2016 {ACM} on {Workshop} on
  {Information} {Sharing} and {Collaborative} {Security}. pp. 35--38. {WISCS}
  '16, ACM, New York, NY, USA (2016). \doi{10.1145/2994539.2994544},
  \url{http://doi.acm.org/10.1145/2994539.2994544}

\bibitem{KENT2015150}
Kent, A.D., Liebrock, L.M., Neil, J.C.: Authentication graphs: Analyzing user
  behavior within an enterprise network. Computers and Security  \textbf{48},
  150 -- 166 (2015). \doi{https://doi.org/10.1016/j.cose.2014.09.001},
  \url{http://www.sciencedirect.com/science/article/pii/S0167404814001321}

\bibitem{kim_lstm-based_2016}
Kim, G., Yi, H., Lee, J., Paek, Y., Yoon, S.: {LSTM}-{Based} {System}-{Call}
  {Language} {Modeling} and {Robust} {Ensemble} {Method} for {Designing}
  {Host}-{Based} {Intrusion} {Detection} {Systems}. arXiv:1611.01726 [cs]  (Nov
  2016), \url{http://arxiv.org/abs/1611.01726}, arXiv: 1611.01726

\bibitem{kolosnjaji_adversarial_2018}
Kolosnjaji, B., Demontis, A., Biggio, B., Maiorca, D., Giacinto, G., Eckert,
  C., Roli, F.: Adversarial {Malware} {Binaries}: {Evading} {Deep} {Learning}
  for {Malware} {Detection} in {Executables}. arXiv:1803.04173 [cs]  (Mar
  2018), \url{http://arxiv.org/abs/1803.04173}, arXiv: 1803.04173

\bibitem{liu_survey_2018}
Liu, Q., Li, P., Zhao, W., Cai, W., Yu, S., Leung, V.C.M.: A {Survey} on
  {Security} {Threats} and {Defensive} {Techniques} of {Machine} {Learning}:
  {A} {Data} {Driven} {View}. IEEE Access  \textbf{6},  12103--12117 (2018).
  \doi{10.1109/ACCESS.2018.2805680}

\bibitem{luo_httpos:_2011}
Luo, X., Zhou, P., Chan, E.W., Lee, W., Chang, R.K., Perdisci, R.: {HTTPOS}:
  {Sealing} {Information} {Leaks} with {Browser}-side {Obfuscation} of
  {Encrypted} {Flows}. In: {NDSS}. vol.~11. Citeseer (2011)

\bibitem{madani_robustness_2018}
Madani, P., Vlajic, N.: Robustness of {Deep} {Autoencoder} in {Intrusion}
  {Detection} {Under} {Adversarial} {Contamination}. In: Proceedings of the 5th
  {Annual} {Symposium} and {Bootcamp} on {Hot} {Topics} in the {Science} of
  {Security}. pp. 1:1--1:8. {HoTSoS} '18, ACM, New York, NY, USA (2018).
  \doi{10.1145/3190619.3190637},
  \url{http://doi.acm.org/10.1145/3190619.3190637}

\bibitem{meier_feedrank:_2018}
Meier, R., Scherrer, C., Gugelmann, D., Lenders, V., Vanbever, L.: {FeedRank}:
  {A} {Tamper}-resistant {Method} for the {Ranking} of {Cyber} {Threat}
  {Intelligence} {Feeds}. In: 2018 10th {International} {Conference} on {Cyber}
  {Conflict} ({CyCon}). IEEE (2018)

\bibitem{mendes_privacy-preserving_2017}
Mendes, R., Vilela, J.P.: Privacy-{Preserving} {Data} {Mining}: {Methods},
  {Metrics}, and {Applications}. IEEE Access  \textbf{5},  10562--10582 (2017).
  \doi{10.1109/ACCESS.2017.2706947}

\bibitem{monaco_what_2019}
Monaco, J.V.: What {Are} {You} {Searching} {For}? {A} {Remote} {Keylogging}
  {Attack} on {Search} {Engine} {Autocomplete}. pp. 959--976 (2019),
  \url{https://www.usenix.org/conference/usenixsecurity19/presentation/monaco}

\bibitem{papernot_limitations_2016}
Papernot, N., McDaniel, P., Jha, S., Fredrikson, M., Celik, Z.B., Swami, A.:
  The {Limitations} of {Deep} {Learning} in {Adversarial} {Settings}. In: 2016
  {IEEE} {European} {Symposium} on {Security} and {Privacy} ({EuroS} {P}). pp.
  372--387 (Mar 2016). \doi{10.1109/EuroSP.2016.36}

\bibitem{Papernot:2017:PBA:3052973.3053009}
Papernot, N., McDaniel, P., Goodfellow, I., Jha, S., Celik, Z.B., Swami, A.:
  Practical black-box attacks against machine learning. In: Proceedings of the
  2017 ACM on Asia Conference on Computer and Communications Security. pp.
  506--519. ASIA CCS '17, ACM, New York, NY, USA (2017).
  \doi{10.1145/3052973.3053009},
  \url{http://doi.acm.org/10.1145/3052973.3053009}

\bibitem{ren_learning-based_2016}
Ren, X., Blanton, R.D., Tavares, V.G.: A {Learning}-{Based} {Approach} to
  {Secure} {JTAG} {Against} {Unseen} {Scan}-{Based} {Attacks}. In: 2016 {IEEE}
  {Computer} {Society} {Annual} {Symposium} on {VLSI} ({ISVLSI}). pp. 541--546
  (Jul 2016). \doi{10.1109/ISVLSI.2016.107}

\bibitem{rigaki_bringing_2018}
Rigaki, M., Garcia, S.: Bringing a {GAN} to a {Knife}-{Fight}: {Adapting}
  {Malware} {Communication} to {Avoid} {Detection}. In: 2018 {IEEE} {Security}
  and {Privacy} {Workshops} ({SPW}). pp. 70--75 (May 2018).
  \doi{10.1109/SPW.2018.00019}

\bibitem{russell_artificial_2009}
Russell, S.J., Norvig, P.: Artificial intelligence: a modern approach. Pearson
  Hall (2009)

\bibitem{sahoo_malicious_2017}
Sahoo, D., Liu, C., Hoi, S.C.H.: Malicious {URL} {Detection} using {Machine}
  {Learning}: {A} {Survey}. arXiv:1701.07179 [cs]  (Jan 2017),
  \url{http://arxiv.org/abs/1701.07179}, arXiv: 1701.07179

\bibitem{shi_benchmarking_2016}
Shi, S., Wang, Q., Xu, P., Chu, X.: Benchmarking {State}-of-the-{Art} {Deep}
  {Learning} {Software} {Tools}. In: 2016 7th {International} {Conference} on
  {Cloud} {Computing} and {Big} {Data} ({CCBD}). pp. 99--104 (Nov 2016).
  \doi{10.1109/CCBD.2016.029}

\bibitem{silver_mastering_2016}
Silver, D., et~al.: Mastering the game of {Go} with deep neural networks and
  tree search. Nature  \textbf{529}(7587),  484--489 (Jan 2016)

\bibitem{sommer_outside_2010}
Sommer, R., Paxson, V.: Outside the {Closed} {World}: {On} {Using} {Machine}
  {Learning} for {Network} {Intrusion} {Detection}. In: 2010 {IEEE} {Symposium}
  on {Security} and {Privacy}. pp. 305--316 (May 2010).
  \doi{10.1109/SP.2010.25}

\bibitem{stoica_berkeley_2017}
Stoica, I., et~al.: A {Berkeley} {View} of {Systems} {Challenges} for {AI}.
  arXiv:1712.05855 [cs]  (Dec 2017), \url{http://arxiv.org/abs/1712.05855}

\bibitem{sun_using_2018}
Sun, X., Dai, J., Liu, P., Singhal, A., Yen, J.: Using {Bayesian} {Networks}
  for {Probabilistic} {Identification} of {Zero}-{Day} {Attack} {Paths}. IEEE
  Transactions on Information Forensics and Security  \textbf{13}(10),
  2506--2521 (Oct 2018). \doi{10.1109/TIFS.2018.2821095}

\bibitem{taylor2016alignment}
Taylor, J., Yudkowsky, E., LaVictoire, P., Critch, A.: Alignment for advanced
  machine learning systems. Machine Intelligence Research Institute  (2016)

\bibitem{theis_note_2015}
Theis, L., Oord, A.v.d., Bethge, M.: A note on the evaluation of generative
  models. arXiv:1511.01844 [cs, stat]  (Nov 2015),
  \url{http://arxiv.org/abs/1511.01844}, arXiv: 1511.01844

\bibitem{tsai_intrusion_2009}
Tsai, C.F., Hsu, Y.F., Lin, C.Y., Lin, W.Y.: Intrusion detection by machine
  learning: {A} review. Expert Systems with Applications  \textbf{36}(10),
  11994--12000 (Dec 2009). \doi{10.1016/j.eswa.2009.05.029},
  \url{http://www.sciencedirect.com/science/article/pii/S0957417409004801}

\bibitem{ucci_survey_2017}
Ucci, D., Aniello, L., Baldoni, R.: Survey on the {Usage} of {Machine}
  {Learning} {Techniques} for {Malware} {Analysis}. arXiv:1710.08189 [cs]  (Oct
  2017), \url{http://arxiv.org/abs/1710.08189}, arXiv: 1710.08189

\bibitem{vasilomanolakis_taxonomy_2015}
Vasilomanolakis, E., Karuppayah, S., MÃŒhlhÃ€user, M., Fischer, M.:
  Taxonomy and {Survey} of {Collaborative} {Intrusion} {Detection}. ACM Comput.
  Surv.  \textbf{47}(4). \doi{10.1145/2716260},
  \url{http://doi.acm.org/10.1145/2716260}

\bibitem{wang_great_2018}
Wang, B., Yao, Y., Viswanath, B., Zheng, H., Zhao, B.Y.: With {Great}
  {Training} {Comes} {Great} {Vulnerability}: {Practical} {Attacks} against
  {Transfer} {Learning}. pp. 1281--1297 (2018),
  \url{https://www.usenix.org/node/217483}

\bibitem{woods_data_2015}
Woods, B., Perl, S.J., Lindauer, B.: Data {Mining} for {Efficient}
  {Collaborative} {Information} {Discovery}. In: Proceedings of the 2Nd {ACM}
  {Workshop} on {Information} {Sharing} and {Collaborative} {Security}. pp.
  3--12. {WISCS} '15, ACM, New York, NY, USA (2015).
  \doi{10.1145/2808128.2808130},
  \url{http://doi.acm.org/10.1145/2808128.2808130}

\bibitem{xu_first_2018}
Xu, M., Liu, J., Liu, Y., Lin, F.X., Liu, Y., Liu, X.: A {First} {Look} at
  {Deep} {Learning} {Apps} on {Smartphones}. arXiv:1812.05448 [cs]  (Nov 2018),
  \url{http://arxiv.org/abs/1812.05448}, arXiv: 1812.05448

\bibitem{yan_cache_2018}
Yan, M., Fletcher, C., Torrellas, J.: Cache {Telepathy}: {Leveraging} {Shared}
  {Resource} {Attacks} to {Learn} {DNN} {Architectures}. arXiv:1808.04761 [cs]
  (Aug 2018), \url{http://arxiv.org/abs/1808.04761}, arXiv: 1808.04761

\bibitem{yue-hei_ng_beyond_2015}
Yue-Hei~Ng, J., Hausknecht, M., Vijayanarasimhan, S., Vinyals, O., Monga, R.,
  Toderici, G.: Beyond {Short} {Snippets}: {Deep} {Networks} for {Video}
  {Classification}. pp. 4694--4702 (2015)

\bibitem{zeiler_visualizing_2013}
Zeiler, M.D., Fergus, R.: Visualizing and {Understanding} {Convolutional}
  {Networks}. arXiv:1311.2901 [cs]  (Nov 2013),
  \url{http://arxiv.org/abs/1311.2901}, arXiv: 1311.2901

\bibitem{zhang2001hide}
Zhang, Z., Li, J., Manikopoulos, C., Jorgenson, J., Ucles, J.: Hide: a
  hierarchical network intrusion detection system using statistical
  preprocessing and neural network classification. In: Proc. IEEE Workshop on
  Information Assurance and Security. pp. 85--90 (2001)

\end{thebibliography}

\end{document}